\documentclass[12pt,preprint]{aastex62}

\usepackage{amssymb}
\usepackage{amsmath}
\usepackage{epstopdf}
\usepackage{xcolor}
\usepackage{amsmath}
\usepackage{graphicx}
\usepackage{hyperref}
\usepackage{graphics,graphicx,natbib}

\def\lapp{\ifmmode\stackrel{<}{_{\sim}}\else$\stackrel{<}{_{\sim}}$\fi}
\def\gapp{\ifmmode\stackrel{>}{_{\sim}}\else$\stackrel{>}{_{\sim}}$\fi}

\newcommand{\degrees}{^{\circ}}

\begin{document}
\title{CHIME/FRB Detection of the Original Repeating Fast Radio Burst Source FRB 121102}
\shorttitle{}
\shortauthors{}

\author{A.~Josephy}
  \affiliation{Department of Physics, McGill University, 3600 rue University, Montr\'eal, QC H3A 2T8, Canada}
  \affiliation{McGill Space Institute, McGill University, 3550 rue University, Montr\'eal, QC H3A 2A7, Canada}
\author{P.~Chawla}
  \affiliation{Department of Physics, McGill University, 3600 rue University, Montr\'eal, QC H3A 2T8, Canada}
  \affiliation{McGill Space Institute, McGill University, 3550 rue University, Montr\'eal, QC H3A 2A7, Canada}
\author{E.~Fonseca}
  \affiliation{Department of Physics, McGill University, 3600 rue University, Montr\'eal, QC H3A 2T8, Canada}
  \affiliation{McGill Space Institute, McGill University, 3550 rue University, Montr\'eal, QC H3A 2A7, Canada}
\author{C.~Ng}
  \affiliation{Dunlap Institute for Astronomy and Astrophysics, University of Toronto, 50 St.~George Street, Toronto, ON M5S 3H4, Canada}
\author{C.~Patel}
  \affiliation{Department of Physics, McGill University, 3600 rue University, Montr\'eal, QC H3A 2T8, Canada}
  \affiliation{McGill Space Institute, McGill University, 3550 rue University, Montr\'eal, QC H3A 2A7, Canada}
\author{Z.~Pleunis}
  \affiliation{Department of Physics, McGill University, 3600 rue University, Montr\'eal, QC H3A 2T8, Canada}
  \affiliation{McGill Space Institute, McGill University, 3550 rue University, Montr\'eal, QC H3A 2A7, Canada}
\author{P.~Scholz}
  \affiliation{Dominion Radio Astrophysical Observatory, Herzberg Astronomy \& Astrophysics Research Centre, National Reseach Council of Canada, P.O.~Box 248, Penticton, BC V2A 6J9, Canada}
\author{B.~C.~Andersen}
  \affiliation{Department of Physics, McGill University, 3600 rue University, Montr\'eal, QC H3A 2T8, Canada}
  \affiliation{McGill Space Institute, McGill University, 3550 rue University, Montr\'eal, QC H3A 2A7, Canada}
\author{K.~Bandura}
  \affiliation{CSEE, West Virginia University, Morgantown, WV 26505, USA}
  \affiliation{Center for Gravitational Waves and Cosmology, West Virginia University, Morgantown, WV 26505, USA}
\author{M.~Bhardwaj}
  \affiliation{Department of Physics, McGill University, 3600 rue University, Montr\'eal, QC H3A 2T8, Canada}
  \affiliation{McGill Space Institute, McGill University, 3550 rue University, Montr\'eal, QC H3A 2A7, Canada}
\author{M.~M.~Boyce}
  \affiliation{Department of Physics and Astronomy, University of Manitoba, Allen Building, Winnipeg, MB R3T 2N2, Canada}
\author{P.~J.~Boyle }
  \affiliation{Department of Physics, McGill University, 3600 rue University, Montr\'eal, QC H3A 2T8, Canada}
  \affiliation{McGill Space Institute, McGill University, 3550 rue University, Montr\'eal, QC H3A 2A7, Canada}
\author{C.~Brar}
  \affiliation{Department of Physics, McGill University, 3600 rue University, Montr\'eal, QC H3A 2T8, Canada}
  \affiliation{McGill Space Institute, McGill University, 3550 rue University, Montr\'eal, QC H3A 2A7, Canada}
\author{D.~Cubranic}
  \affiliation{Department of Physics \& Astronomy, University of British Columbia, 6224 Agricultural Road, Vancouver, BC V6T 1Z1, Canada}
\author{M.~Dobbs}
  \affiliation{Department of Physics, McGill University, 3600 rue University, Montr\'eal, QC H3A 2T8, Canada}
  \affiliation{McGill Space Institute, McGill University, 3550 rue University, Montr\'eal, QC H3A 2A7, Canada}
\author{B.~M.~Gaensler }
  \affiliation{Dunlap Institute for Astronomy and Astrophysics, University of Toronto, 50 St.~George Street, Toronto, ON M5S 3H4, Canada}
  \affiliation{Department of Astronomy and Astrophysics, University of Toronto, 50 St.~George Street, Toronto, ON M5S 3H4, Canada}
\author{A.~Gill}
  \affiliation{Dunlap Institute for Astronomy and Astrophysics, University of Toronto, 50 St.~George Street, Toronto, ON M5S 3H4, Canada}
  \affiliation{Department of Astronomy and Astrophysics, University of Toronto, 50 St.~George Street, Toronto, ON M5S 3H4, Canada}
\author{U.~Giri}
  \affiliation{Perimeter Institute for Theoretical Physics, 31 Caroline Street N, Waterloo, ON N2L 2Y5, Canada}
  \affiliation{Department of Physics and Astronomy, University of Waterloo, Waterloo, ON N2L 3G1, Canada}
\author{D.~C.~Good}
  \affiliation{Department of Physics \& Astronomy, University of British Columbia, 6224 Agricultural Road, Vancouver, BC V6T 1Z1, Canada}
\author{M.~Halpern}
  \affiliation{Department of Physics and Astronomy, University of British Columbia, 6224 Agricultural Road, Vancouver, BC V6T 1Z1, Canada}
\author{G.~Hinshaw}
  \affiliation{Department of Physics and Astronomy, University of British Columbia, 6224 Agricultural Road, Vancouver, BC V6T 1Z1, Canada}
\author{V.~M.~Kaspi}
  \affiliation{Department of Physics, McGill University, 3600 rue University, Montr\'eal, QC H3A 2T8, Canada}
  \affiliation{McGill Space Institute, McGill University, 3550 rue University, Montr\'eal, QC H3A 2A7, Canada}
\author{T.~L.~Landecker}
  \affiliation{Dominion Radio Astrophysical Observatory, Herzberg Astronomy \& Astrophysics Research Centre, National Reseach Council of Canada, P.O.~Box 248, Penticton, BC V2A 6J9, Canada}
\author{D.~A.~Lang}
  \affiliation{Perimeter Institute for Theoretical Physics, 31 Caroline Street N, Waterloo, ON N2L 2Y5, Canada}
\author{H.-H.~Lin}
  \affiliation{Canadian Institute for Theoretical Astrophysics, University of Toronto, 60 St.~George Street, Toronto, ON M5S 3H8, Canada}
\author{K.~W.~Masui}
  \affiliation{MIT Kavli Institute for Astrophysics and Space Research, Massachusetts Institute of Technology, 77 Massachusetts Ave, Cambridge, MA 02139, USA}
  \affiliation{Department of Physics, Massachusetts Institute of Technology, 77 Massachusetts Ave, Cambridge, MA 02139, USA}
\author{R.~Mckinven}
  \affiliation{Dunlap Institute for Astronomy and Astrophysics, University of Toronto, 50 St.~George Street, Toronto, ON M5S 3H4, Canada}
  \affiliation{Department of Astronomy and Astrophysics, University of Toronto, 50 St.~George Street, Toronto, ON M5S 3H4, Canada}
\author{J.~Mena-Parra}
  \affiliation{MIT Kavli Institute for Astrophysics and Space Research, Massachusetts Institute of Technology, 77 Massachusetts Ave, Cambridge, MA 02139, USA}
\author{M.~Merryfield}
  \affiliation{Department of Physics, McGill University, 3600 rue University, Montr\'eal, QC H3A 2T8, Canada}
  \affiliation{McGill Space Institute, McGill University, 3550 rue University, Montr\'eal, QC H3A 2A7, Canada}
\author{D.~Michilli}
  \affiliation{Department of Physics, McGill University, 3600 rue University, Montr\'eal, QC H3A 2T8, Canada}
  \affiliation{McGill Space Institute, McGill University, 3550 rue University, Montr\'eal, QC H3A 2A7, Canada}
\author{N.~Milutinovic}
  \affiliation{Dominion Radio Astrophysical Observatory, Herzberg Astronomy \& Astrophysics Research Centre, National Reseach Council of Canada, P.O.~Box 248, Penticton, BC V2A 6J9, Canada}
  \affiliation{Department of Physics and Astronomy, University of British Columbia, 6224 Agricultural Road, Vancouver, BC V6T 1Z1, Canada}
\author{A.~Naidu}
  \affiliation{Department of Physics, McGill University, 3600 rue University, Montr\'eal, QC H3A 2T8, Canada}
  \affiliation{McGill Space Institute, McGill University, 3550 rue University, Montr\'eal, QC H3A 2A7, Canada}
\author{U.~Pen}
  \affiliation{Canadian Institute for Theoretical Astrophysics, University of Toronto, 60 St.~George Street, Toronto, ON M5S 3H8, Canada}
\author{M.~Rafiei-Ravandi}
  \affiliation{Perimeter Institute for Theoretical Physics, 31 Caroline Street N, Waterloo, ON N2L 2Y5, Canada}
\author{M.~Rahman}
  \affiliation{Dunlap Institute for Astronomy and Astrophysics, University of Toronto, 50 St.~George Street, Toronto, ON M5S 3H4, Canada}
\author{S.~M.~Ransom}
  \affiliation{NRAO, 520 Edgemont Rd., Charlottesville, VA 22903, USA}
\author{A.~Renard}
  \affiliation{Dunlap Institute for Astronomy and Astrophysics, University of Toronto, 50 St.~George Street, Toronto, ON M5S 3H4, Canada}
\author{S.~R.~Siegel}
  \affiliation{Department of Physics, McGill University, 3600 rue University, Montr\'eal, QC H3A 2T8, Canada}
  \affiliation{McGill Space Institute, McGill University, 3550 rue University, Montr\'eal, QC H3A 2A7, Canada}
\author{K.~M.~Smith}
  \affiliation{Perimeter Institute for Theoretical Physics, 31 Caroline Street N, Waterloo, ON N2L 2Y5, Canada}
\author{I.~H.~Stairs}
  \affiliation{Department of Physics \& Astronomy, University of British Columbia, 6224 Agricultural Road, Vancouver, BC V6T 1Z1, Canada}
\author{S.~P.~Tendulkar}
  \affiliation{Department of Physics, McGill University, 3600 rue University, Montr\'eal, QC H3A 2T8, Canada}
  \affiliation{McGill Space Institute, McGill University, 3550 rue University, Montr\'eal, QC H3A 2A7, Canada}
\author{K.~Vanderlinde}
  \affiliation{Dunlap Institute for Astronomy and Astrophysics, University of Toronto, 50 St.~George Street, Toronto, ON M5S 3H4, Canada}
  \affiliation{Department of Astronomy and Astrophysics, University of Toronto, 50 St.~George Street, Toronto, ON M5S 3H4, Canada}
\author{P.~Yadav}
  \affiliation{Department of Physics \& Astronomy, University of British Columbia, 6224 Agricultural Road, Vancouver, BC V6T 1Z1, Canada}
\author{A.~V.~Zwaniga}
  \affiliation{Department of Physics, McGill University, 3600 rue University, Montr\'eal, QC H3A 2T8, Canada}
  \affiliation{McGill Space Institute, McGill University, 3550 rue University, Montr\'eal, QC H3A 2A7, Canada}

\accepted{2019 July 23}
\submitjournal{\apjl}

\begin{abstract}
We report the detection of a single burst from the first-discovered repeating Fast Radio Burst source, FRB 121102, with CHIME/FRB, which operates in the frequency band 400--800 MHz.  The detected burst occurred on 2018 November 19 and its emission extends down to at least 600~MHz, the lowest frequency detection of this source yet. The burst, detected with a significance of $23.7\sigma$, has fluence $12\pm3$~Jy~ms and shows complex time and frequency morphology. The 34~ms width of the burst is the largest seen for this object at any frequency. We find evidence of sub-burst structure that drifts downward in frequency at a rate of $-3.9\pm0.2$~MHz~ms$^{-1}$. Our best fit tentatively suggests a dispersion measure of $563.6\pm0.5$~pc~cm$^{-3}$, which is ${\approx}1$\% higher than previously measured values. We set an upper limit on the scattering time at 500~MHz of 9.6~ms, which is consistent with expectations from the extrapolation from higher frequency data. We have exposure to the position of FRB 121102 for a total of 11.3~hrs within the FWHM of the synthesized beams at 600~MHz from 2018 July~25 to 2019 February~25. We estimate on the basis of this single event an average burst rate for FRB 121102 of 0.1--10 per day in the 400--800~MHz band for a median fluence threshold of 7~Jy~ms in the stated time interval.
\end{abstract}

%\keywords{pulsars: general --- pulsars: timing --- surveys, RRATs}
%\keywords{radio transient sources --- radio bursts --- surveys}

\section{Introduction}

Fast Radio Bursts (FRBs) are a recently recognized astrophysical phenomenon \citep{lbm+07} consisting of short (few millisecond) bursts of radio waves coming from apparently cosmological distances.  Their physical nature is as yet unknown.  

A major clue to the FRB puzzle came with the discovery of the first repeating source, FRB 121102 \citep{sch+14,ssh+16a}.  Its repeating nature enabled its interferometric localization and the identification of a host dwarf galaxy at redshift $z=0.2$ \citep{clw+17,tbc+17}. FRB 121102 is also remarkable for its sometimes complex burst phenomenology \citep{ssh+16b,hss+18} which involves highly variable spectra and sub-bursts that appear to drift downward in radio frequency with time. The repetition also allows studies of burst properties in different wavebands. Until now, all such observations have been carried out at radio frequencies above 1 GHz, with bursts detected as high as 8 GHz \citep{msh+18,gsp+18}.

Of key interest is the environment around FRB 121102.  \citet{msh+18} showed the source has an exceptionally high Faraday rotation measure (${\sim}10^5$ rad m$^{-2}$) indicating extreme magneto-ionic surroundings.  This plus a persistent, variable continuum radio source co-located with FRB 121102 \citep{clw+17,mph+17} has inspired a model involving a young magnetar in an expanding supernova remnant \citep{mbm17,mm18,mms19} that potentially explains many of the observed properties. 

One important property is the source's scattering time, which has imprinted information on the distribution of ionized matter along the line of sight.  Until now, the bursts have shown no evidence of the expected temporal asymmetry associated with multi-path scattering, with an upper limit on pulse broadening of 1.5~ms at 1.5~GHz \citep{ssh+16a}. On the other hand, observations at 5~GHz have measured a scintillation bandwidth consistent with a scatter-broadening time of 24~$\mu$s at 1~GHz \citep{msh+18}.  So low a scattering time suggests that bursts from FRB 121102 may be detectable in the LOFAR band, despite no bursts having been seen thus far \citep{hst+19}. However, \citet{mls+15a}, for a different FRB, measured a scatter-broadening time near 800~MHz inconsistent with that inferred from the event's scintillation bandwidth, i.e.\ two scattering time scales for one FRB. This demonstrates the existence of two distinct scattering screens toward that source, a situation that could also be true for FRB 121102.  A measurement of scatter broadening would therefore be interesting, but requires a detection below 1~GHz.

Here we report the detection of a single burst from FRB 121102 in the frequency
range 400--800 MHz using the Canadian Hydrogen Intensity Mapping Experiment (CHIME)
telescope and its Fast Radio Burst (FRB) detection system \citep{abb+18}.  We also report
on data from the CHIME/Pulsar instrument
%(CHIME/Pulsar Collaboration, in prep.)
, obtained simultaneously with CHIME/FRB, in which the burst also appears.  

\section{Observations and Analysis}
\label{sec:obs}

\begin{figure}[t]
	\centering
\includegraphics[scale=0.7]{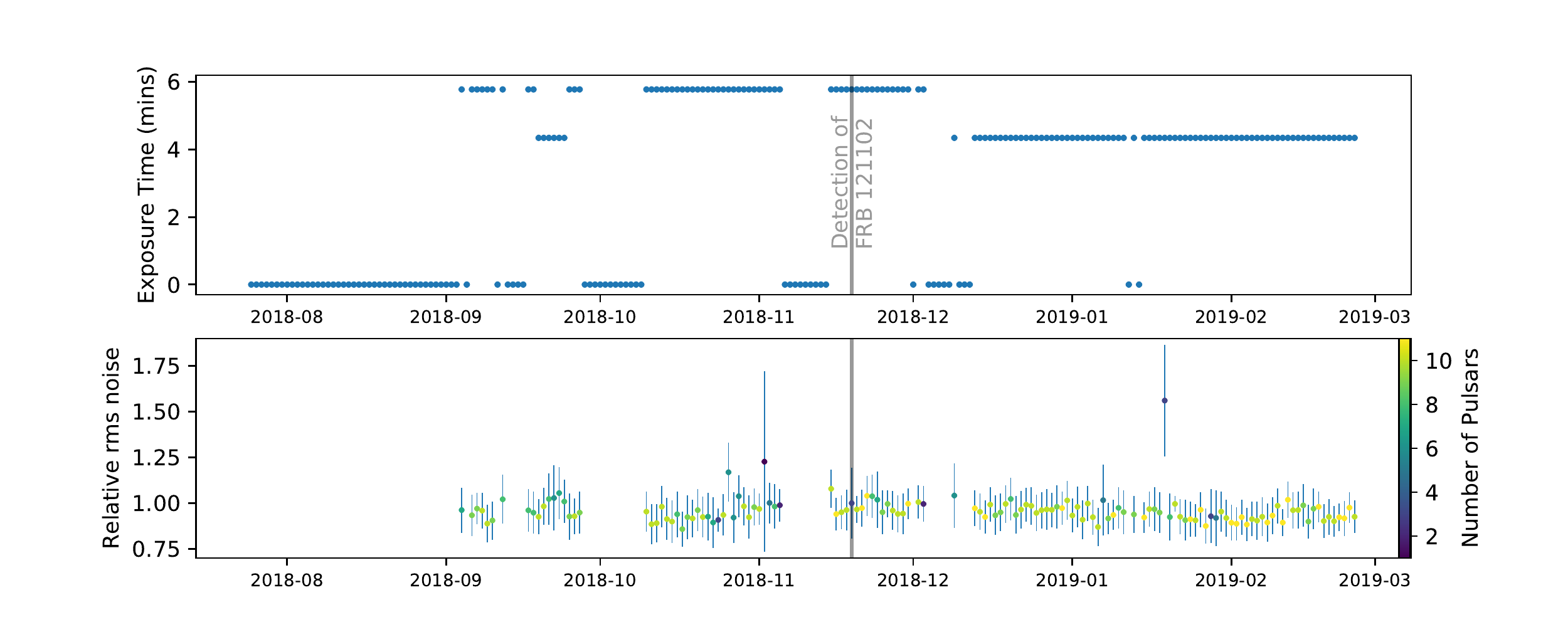} %{rel_sensitivity_observing_time_R1.pdf}
\figcaption{Timeline of CHIME/FRB's daily exposure to FRB 121102 within the
FWHM of the telescope's synthesized beams at 600~MHz.
%telescope's primary beam.
The reduction in daily exposure starting 2018 December 7 is due to failure of the computing node responsible for processing one of the four FFT-formed beams through which the source transits. 
The RMS noise is estimated relative to that for the day of detection of the FRB 121102 burst using pulsars detected by CHIME/FRB with the marker colors denoting the number of pulsars used for the estimate. The errors on the relative RMS noise account only for the day-to-day sensitivity variation and are not representative of the overall error on the fluence threshold. See \S\ref{sec:sensitivity} for a description of how the fluence threshold and the corresponding error were determined.}
\label{fig:exposure}
\end{figure}

\subsection{CHIME/FRB Detection}
\label{sec:frbdetection}

The CHIME/FRB instrument has been described in an overview paper in which the CHIME telescope, FRB detection instrument, and pipeline are described in detail \citep{abb+18}.  During the interval from 2018 July 25 to 2019 February 25, CHIME was in a commissioning state in which various components of the instrument were being tested, with software and calibration systems being updated and improved frequently. Although the CHIME/FRB system was operational starting 2018 July 25, the beam configuration for the months of July and August rendered FRB 121102 undetectable as it did not transit within the FWHM region of any of the FFT-formed beams at 600 MHz. Following a system reconfiguration on 2018 September 4, we were sensitive to FRB 121102 for a total of 11.3~hours, as shown in Figure~\ref{fig:exposure}. We truncate our reported exposure on 2019 February 25, when we brought the CHIME/FRB system down for pipeline updates and testing.
The telescope sensitivity was known to be varying from day to day during the interval due primarily to changing gain calibration strategies, but also due to a variety of issues that,
once recognized, were rectified.  Overall, we determine the median sensitivity to be $7\,^{+ 7}_{- 4}$~Jy~ms, for a burst width of 34~ms (see \S\ref{sec:morphology}), during on times shown in Figure~\ref{fig:exposure}. Further details regarding the sensitivity estimate are provided below (see \S\ref{sec:sensitivity}). 

\begin{figure}[h]
	\centering
\includegraphics[scale=0.5]{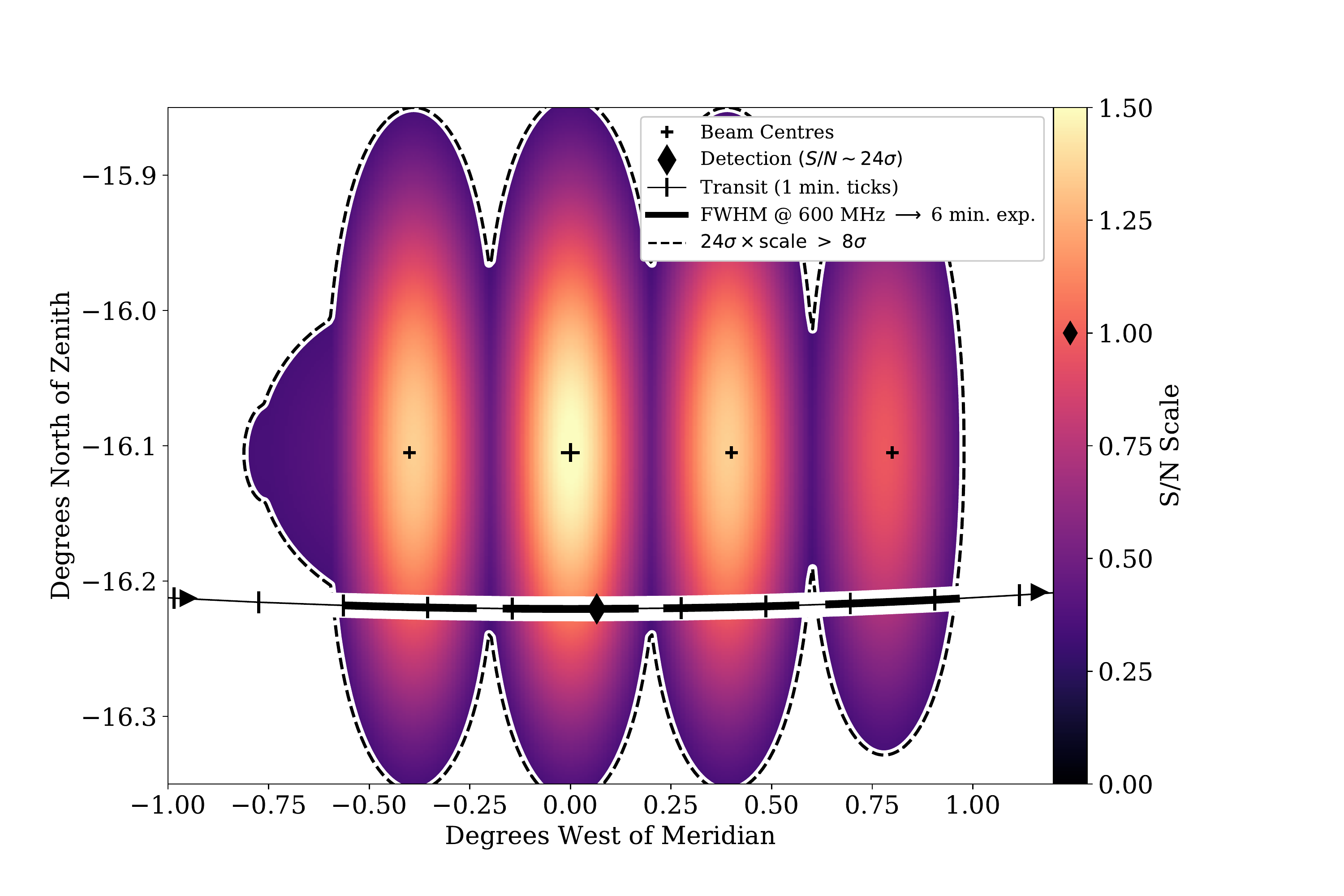} %{transit_sensitivity.pdf}
\figcaption{Detection location and transit sensitivity for FRB 121102. The color scale reflects sensitivity variation in the four beams relevant for detection.  Crosses indicate beam centers, with the largest cross denoting the detection beam center. The true location at the time of burst detection is indicated with a diamond. Protrusions on the left are caused by east-west aliasing of the synthesized beams on the right. The intersection of the transit with the FWHM at 600 MHz for each beam (indicated with thick black lines) is used for exposure calculations.}
\label{fig:localization}
\end{figure}

The burst was detected and associated with FRB 121102 by our automated FRB-detection pipeline on 2018 November 19, at 09:38:47.706 UTC (topocentric, 600~MHz), when our threshold for saving buffered intensity data was signal-to-noise ratio, S/N=10.  The nominal pipeline detection S/N for the event was 23.7, and so intensity data were saved.  The event was detected in a single beam whose central pointing position is consistent with the known position of FRB 121102 \citep{clw+17}, as shown in Figure~\ref{fig:localization}.  This, and the similarity in nominal pipeline DM of 565~pc\,cm$^{-3}$ to previously published values \cite[${\sim}559$~pc\,cm$^{-3}$; ][]{ssh+16b}, assured us of the event's identification.

\begin{figure}[h]
	\centering
\includegraphics[scale=0.5]{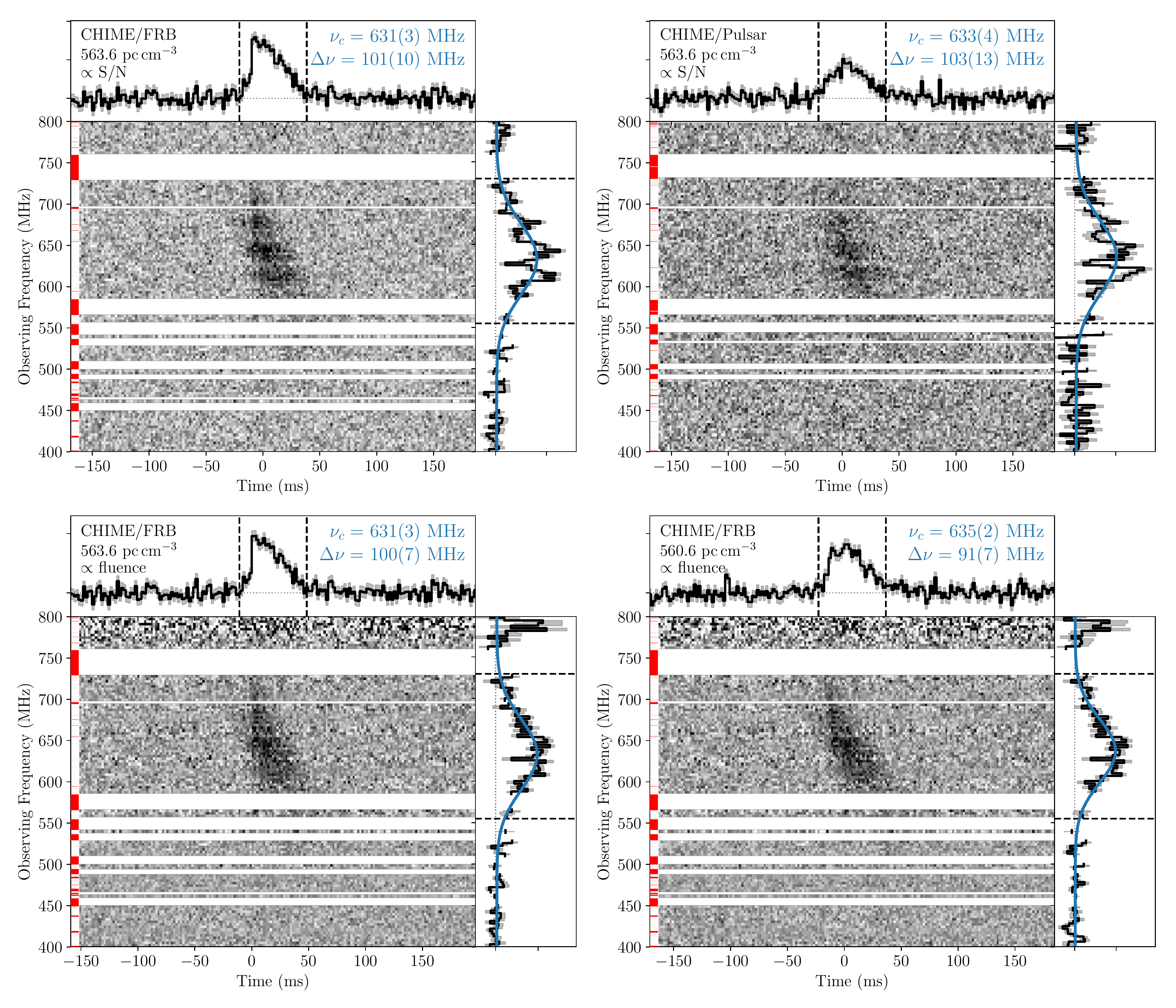} %{waterfalls.pdf}
\figcaption{
Signal intensity as a function of frequency and time (``waterfall'') plot of the CHIME/FRB (upper left and lower two) and CHIME/Pulsar (upper right) detections of FRB 121102. Data have been dedispersed to the structure-optimizing DM of 563.6~pc~cm$^{-3}$, except for the lower right subplot, which is dedispersed to the nominal September 2016 value of 560.6~pc~cm$^{-3}$ \citep{ssh+16a}. Data plotted here are downsampled to a frequency resolution of ${\sim}$3~MHz and a time resolution of ${\sim}$2~ms. For the upper subplots, the greyscale reflects S/N of the source dynamic spectrum without explicit bandpass correction. Each channel is independently normalized according to statistics of the off-pulse regions (delineated by vertical dashed lines in top panel). In the lower subplots, bandpass calibration has been applied, and the greyscale reflects fluence. Note that bandpass correction is a work in progress and is complicated  by multiple factors --- see text for details. The red lines at the left sides of plots represent radio frequencies masked prior to analysis, while horizontal white stripes show the full extent of frequencies removed after all interference rejection.  The right panel is the projected on-pulse spectrum. Strong (order 1) instrumental variations in bandpass are visible in the uncorrected subplots. The blue line is the best single-Gaussian-fit to the spectrum (annotated with center frequency and FWHM). The top panel contains the total pulse profile after summing over the frequency channels that bracket the burst (delineated with horizontal dashed lines in right panel; 550$-$730~MHz).}
\label{fig:waterfall}
\end{figure}

The saved intensity data, having 16k frequency channels each at 0.983-ms time resolution, allowed us to produce a ``waterfall plot'', as shown in Figure~\ref{fig:waterfall}.  A complex burst morphology is seen, both in time and in radio frequency.  Note that correcting for CHIME's bandpass response is a work in progress, and is complicated by a declination-dependent $\sim$30-MHz ripple due to multiple reflections, declination-specific ${\sim}$7-MHz ripple due to FFT beamforming \citep{nvp+17}, as well as additional rippling due to the use of a polyphase filterbank algorithm and FFT upchannelization for defining our frequency channels \citep{abb+18}.  Bandpass-calibrated data are shown in the bottom subplots of Figure~\ref{fig:waterfall}.  The conversion from instrumental units to fluence is discussed in \S\ref{sec:fluence}.
%, and a detailed description on how the bandpass calibration is accomplished will be provided in a future paper.
Emission is clearly detected down to $\sim$600~MHz, by far the lowest yet seen for this source.  Moreover, it also appears to be band-limited, with no emission seen below this value, in spite of good sensitivity down to 400 MHz.

\newpage

\subsection{CHIME/Pulsar Detection}
\label{sec:pulsardetection}

The CHIME X-Engine, in addition to supplying data to the CHIME/FRB instrument as described in \citet{abb+18}, also provides tied-array formed-beam data products to a separate backend that is used for targeted observations of pulsars and other known radio transients.  This instrument, CHIME/Pulsar, is described in detail in \citet{n++17}.
%and CHIME/Pulsar Collaboration, in prep. 
CHIME/Pulsar has been monitoring the position of FRB~121102 reported in \citet{clw+17} since 2018 November 3. Each observation tracks the position of the source for about 19\,min as it drifts over the meridian FOV of CHIME. All together, 23.86\,hr were spent on source, covering a total of 81 days of transits up until 2019 March 1. The data were coherently dedispersed to a DM\footnote{This DM was chosen ahead of time based on previously published values \citep[e.g.][]{ssh+16b}. The difference compared to our newly measured value is sufficiently small that it does not affect the CHIME/Pulsar pulse profile given the narrow channel widths.} of 558.1\,pc\,cm$^{-3}$ then saved as a total-power filterbank time series, with a time resolution of 327.68\,$\mu$s and 1024 frequency channels each with a width of 390\,kHz. Offline, a {\textsc PRESTO}-based \citep{presto} single-pulse search was conducted on these data. Throughout this period, only one burst was observed, on 2018 November 19, for which the CHIME/FRB has a simultaneous detection. The waterfall plot of this burst is shown in Figure~\ref{fig:waterfall}. This burst was detected with a S/N of 17 using the CHIME/Pulsar instrument, and shows temporal structure similar to previous bursts of FRB~121102. We note that the S/N reported from the Pulsar backend is lower than that from the FRB backend. The difference is likely due to imperfect scaling from floating point precision to the 4-bit integer VDIF output in the early commissioning data of the Pulsar backend. For this reason, we restrict further analysis and rate estimates to the FRB backend, where defining a fluence completeness limit throughout our quoted exposure is more tractable.

\subsection{Burst Morphology Analysis}
\label{sec:morphology}

The observed burst morphology --- broad pulse structure with several apparent components --- might be caused by a combination of sub-burst drifting and scattering. To try to disentangle the two, we analyzed the CHIME/FRB intensity data first by assuming all morphology is caused by sub-burst drifting, and then by fitting a model burst that includes both sub-bursts and scattering.

\begin{figure}[h]
	\centering
\includegraphics[scale=0.5]{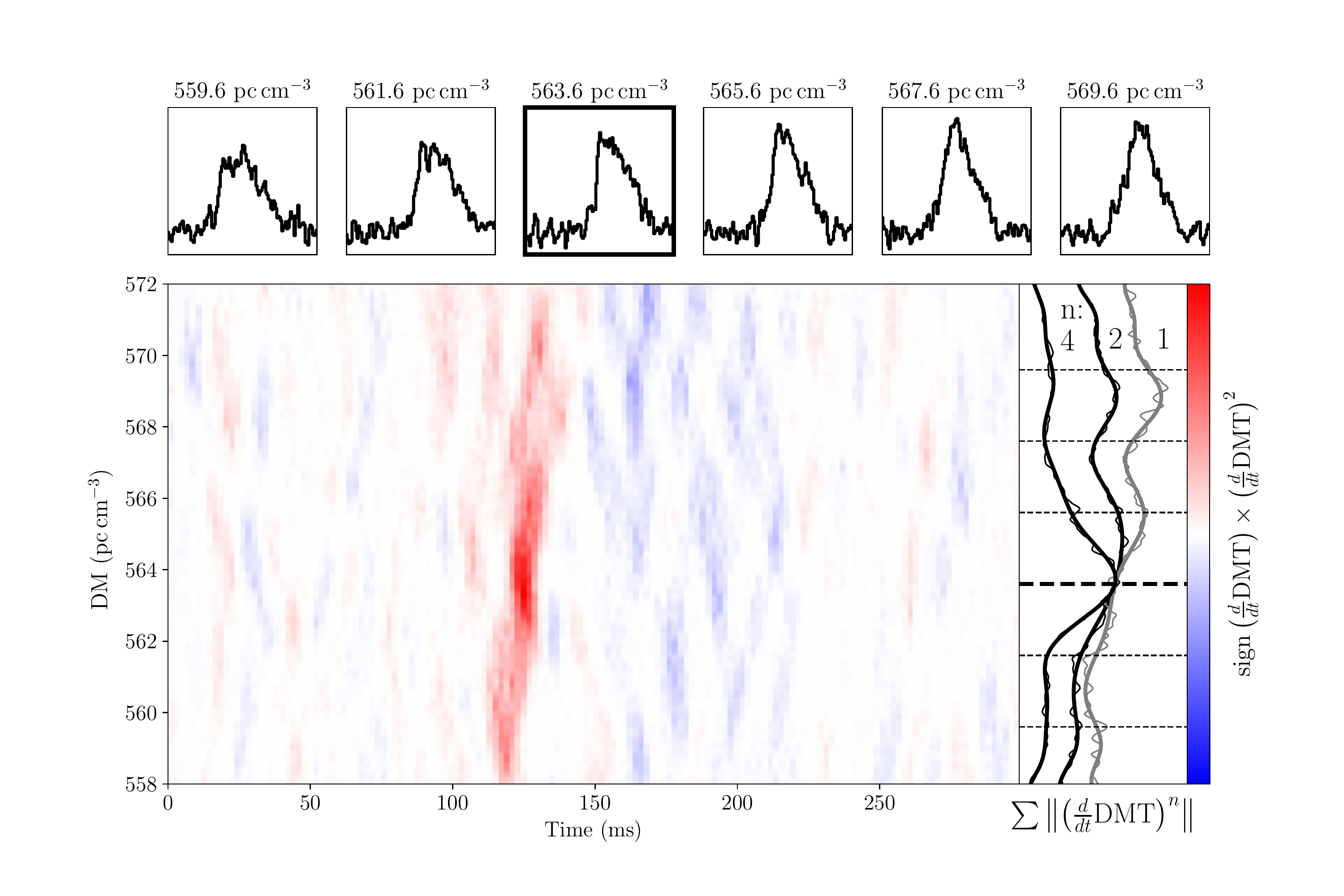} %{structure_optimized_DM.pdf}
\figcaption{Structure-optimizing DM method. The main panel shows the signed square ($n = 2$) of the forward time-derivative of the smoothed DM transform (DMT), where the color scale reflects the local steepness of frequency-averaged burst profiles for different DMs. While absolute values are taken for the final metric, we show the sign to highlight rising and falling regions. Likewise, we show $n=2$ to highlight structure beyond a singular sharp rise. The curves in the right panel are high-order polynomial fits to time-averaged time-derivatives to the power $n$, normalized to DM 563.6~pc~cm$^{-3}$. Dashed lines correspond to the DMs used to produce the six frequency-averaged burst profiles at the top, with the structure optimizing DM indicated in bold.  Note that the profiles have been convolved with a 3-ms boxcar to match the smoothing used on the DM transform prior to taking the time-derivative.}
\label{fig:structure_DM}
\end{figure}

To characterize sub-burst drifting in the burst, we optimized the DM for burst structure by finding the DM that maximizes the forward-derivative of the dedispersed time series, following \citet{hss+18} and \citet{gsp+18}, as shown in Figure~\ref{fig:structure_DM}. We calculated the DM transform for 400 steps ranging from 556 to 574 pc cm$^{-3}$ and smoothed the transform with a $3\times3$ (2.8-ms $\times$ 0.135 pc cm$^{-3}$) rectangular kernel to reduce the noise. Then, we calculated the forward-derivative and smoothed the resulting image further by a $3\times5$ (2.8-ms $\times$ 0.225 pc cm$^{-3}$) rectangular kernel. Finally, we summed the absolute value of the forward-derivative to the power $n \in(1, 2, 4)$ over the time axis. The approach of \citet{gsp+18} corresponds to $n = 1$, while \citet{hss+18} use $n = 2$. We find that higher values of $n$ select for singular sharp rises in the pulse-profile, while lower values may favor multiple lower-amplitude peaks. For a given $n$, the maximum of a high-order polynomial fit to the curve is taken as the structure-optimizing DM. To estimate uncertainties, we normalize the DM transform such that each pixel represents a S/N value, then perform a Monte Carlo simulation by perturbing the transform with normally distributed noise. For $n = 4$, we find a structure-optimizing DM of $563.6\pm0.5$ pc$\,$cm$^{-3}$, while $n = 2$ gives $565\pm1$ pc$\,$cm$^{-3}$. We adopt the former value as the curve for $n = 4$ has an unambiguous peak. This  is several DM units higher than earlier reported DMs for this source \cite[e.g.][]{ssh+16a,hss+18} but is consistent with measurements of the structure-optimizing DM in Arecibo observations near 1.4 GHz from November 2018 (Seymour et al. in prep.).

\begin{figure}[h]
    \centering
\includegraphics[scale=0.5]{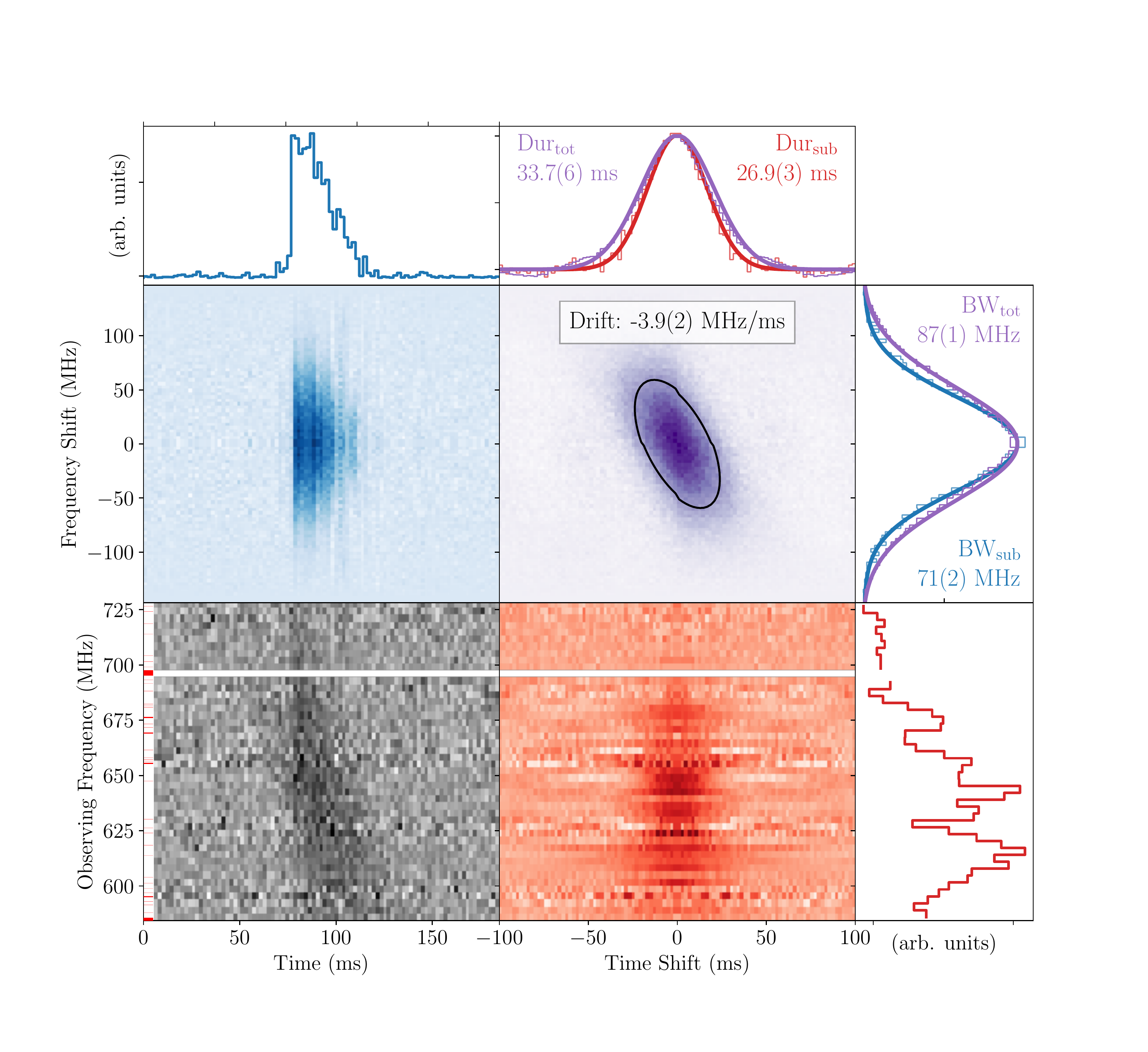} %{autocorrelation_analysis.pdf}
\figcaption{Auto-correlations of the burst emission region. Bottom left (gray): the dedispersed dynamic spectrum with a frequency resolution of ${\sim}$3~MHz and time resolution of ${\sim}$3~ms (corrected for the effective bandpass).  Bottom right (red): the 1D auto-correlation of every time series shows the characteristic width of the sub-bursts for different frequencies. Top left (blue): the 1D auto-correlation of every spectrum shows the characteristic sub-burst bandwidth for different times. Top right (purple): the 2D auto-correlation shows the self-similarity of the burst and the sub-burst drift rate, obtained by fitting a 2D Gaussian (FWHM indicated in black). Summed auto-correlations, with colors corresponding to the images, are shown in the panels on top and at the right. Note that structure in the projected 1D time auto-correlation is largely instrumental.}
\label{fig:autocorrelation}
\end{figure}

Using the structure-optimizing DM, we analyze the auto-correlation of the emission region in the dedispersed dynamic spectrum, as shown in Figure~\ref{fig:autocorrelation}. To mitigate RFI masking effects, we limit the analysis to the relatively clean 580--725~MHz band, which contains the majority of the observed signal (see Fig.~\ref{fig:waterfall}). The 2D auto-correlation shows where the burst has the most self-similarity. A tilt in the ellipse reflects the sub-burst drift rate and here we find a rate $-3.9\pm0.2$ MHz ms$^{-1}$. Assuming the pulse profile and spectral bandwidth of the burst envelope and sub-bursts are reasonably well described by Gaussian profiles, we measure their FWHMs from the standard deviation $\sigma$ of a Gaussian profile fit to the summed auto-correlation over the respective axes\footnote{As the auto-correlation of a Gaussian profile has a FWHM $\sqrt{2}$ times the original width and the FWHM of a Gaussian is $2 \sqrt{2 \ln 2} \sigma$, we have FWHM$_\mathrm{burst}$ = $2 \sqrt{\ln 2} \sigma_\mathrm{corr}$.}. We find a burst envelope width of $33.7\pm0.6$ ms and a typical sub-burst width of $26.9\pm0.3$ ms. The total emission bandwidth is $87\pm1$ MHz and the sub-burst bandwidth $71\pm2$ MHz. We note that bandpass correction is a work in progress, imposes structure on comparable scales, but which is stable on the time scale of the burst. The 530--580 MHz range, excluded due to significant RFI contamination, may hide additional structure and lead to an underestimated envelope bandwidth and duration. This possibility is supported by the Gaussian fit to the full-band spectrum (see Fig.~\ref{fig:waterfall}), which is centred at 631~MHz. Note that due to the limited S/N of the detection, coupled with imperfect bandpass correction, the intrinsic structure of the burst remains mostly unresolved --- despite time resolutions much finer than the envelope width.

In order to quantify frequency-dependent scattering in the burst, we also applied a S/N-optimizing routine to our raw total-intensity data that uses a least-squares algorithm similar to that used in the analysis of the first 13 FRBs discovered by CHIME/FRB \citep{abb+19a}. This algorithm was modified to allow for fitting an arbitrary number of Gaussian spectral components and their respective parameters against 16k-channelized dynamic spectra; per-component parameters include an arrival time, time/frequency gaussian widths, signal amplitude and central frequency of peak emission. The DM and scattering parameters are applied to all sub-structure components as ``global" fit parameters. 

We find that three Gaussian components are statistically favored for FRB 121102 when using the spectrum-fitting algorithm on our bandpass-calibrated data, and weighting spectra residuals by per-channel variances. In these fits, we hold the DM fixed to the structure-optimizing value of 563.6 pc cm$^{-3}$, as well as the DM and scattering indices fixed to $-$2 and $-$4, respectively. We detect no significant scattering of the burst, with a 3-sigma upper limit on the scattering timescale of 9.6~ms at 500~MHz. Fit parameters are listed in Table~\ref{table:fitparams}.

The best-fit parameters of the three Gaussian components can also be used to constrain the sub-burst drift rate in time and frequency. Using these data, we find the drift rate between the three components to be $-2.9\pm1.0$ MHz ms$^{-1}$ when using a orthogonal-distance-regression method for weighting fits by the uncertainties in spectral and temporal centroid positions of the three components. This rate is consistent with the estimate from the model-independent auto-correlation analysis.

We note that the S/N of our data and the complex bandpass function do not allow us to robustly search for evidence of diffractive scintillation in the spectrum.

\begin{table}
    \centering
    \begin{tabular}{lccc}
        \hline 
        Parameter & \multicolumn{3}{c}{Global Parameters} \\
        \hline\hline
        Dispersion Measure (pc cm$^{-3}$) & \multicolumn{3}{c}{563.6} \\
        Dispersion Index & \multicolumn{3}{c}{$-$2} \\
        Scattering Timescale (ms; referenced to 1 GHz) & \multicolumn{3}{c}{0.27(11)} \\
        Scattering Index & \multicolumn{3}{c}{$-$4} \\
        \hline
        Parameter & Component 1 & Component 2 & Component 3 \\
        \hline\hline 
        Arrival time relative to first component (ms) & 0(2) & 8(2) & 27(2) \\
        Amplitude (Jy) & 0.6(2) & 2.4(3) & 0.7(4) \\
        Time Width (ms) & 3.1(5) & 10.1(9) & 7(2) \\
        Frequency Width (MHz) & 26(4) & 33(4) & 18(8) \\
        Frequency of Peak Emission & 684(4) & 644(6) & 612(10) \\
        \hline
    \end{tabular}
    \caption{Best-fit Gaussian components for dynamic-spectrum substructure of FRB 121102. Quoted uncertainties represent 68.3\% confidence intervals. Values that were fixed during model fit are listed without uncertainties.}
    \label{table:fitparams}
\end{table}

\newpage

\subsection{Burst Fluence Determination}
\label{sec:fluence}

The burst was detected during the commissioning phase of the CHIME/FRB system. We therefore adopt the same method to estimate its fluence as in \citet{abb+19a}. We used the observation of 3C~48 (which has a declination within 0.1$^\circ$ of FRB 121102) on 2019 November 18 for calibration and obtained a flux conversion factor as a function of radio frequency to account for the telescope primary beam. We assessed the uncertainty on the fluence in the same way as described in \citet{abb+19a}. Here, we used 9 bright sources within 1$^\circ$ (instead of 5$^\circ$) elevation angle. Since we estimated an uncertainty on the flux as a function of frequency using the calibration sources, we have an upper and lower bound for each intensity value in the band-averaged profile, hence, an upper and a lower bound profile. We measured the uncertainty on the burst fluence by measuring the average area between the two profiles. Since we were operating in a commissioning phase, a large fraction of the bandwidth was unusable for estimating the fluence for the same reasons described in \citet{abb+19a}. Overall, we estimated the burst fluence to be $\rm 12\pm3 $ Jy~ms, measured over an effective total bandwidth of 255 MHz in the range 400--800 MHz.
We stress that uncertainty in our fluence measurement is systematics dominated, and is expected to improve as our beam model and calibration techniques mature. The detection S/N ($23.7\sigma$) should be taken as the best representation of detection significance.

\subsection{System Sensitivity Estimate}
\label{sec:sensitivity}

In determining sensitivity to FRB 121102 over the course of our exposure, we characterize three potential sources of variability:  (i) day-to-day instrument gain variations, (ii) changing source position within the synthesized beams, and (iii) different emission bandwidths and frequency centers within the instrument bandpass.

To capture (i), the day-to-day gain variation, we identified Galactic radio pulsars within $5\degrees$ declination of the source, which were detected by CHIME/FRB on at least 20\% of the days between 2018 July 25 and 2019 February 25. For the purpose of this analysis, we identify a pulsar detection as robust if at least 5 pulses with S/N $>$ 8 were observed on each day within the FWHM region for the FFT-formed beams at 600 MHz. We use the distribution of measured S/N values for these pulses to estimate the RMS radiometer noise, $T_\mathrm{sys}/(G \sqrt{n_\mathrm{p} \Delta \nu t_\mathrm{samp}})$, where $T_\mathrm{sys}$ is the system temperature, $G$ is the telescope gain, $n_\mathrm{p}$ is the number of summed polarizations, $\Delta \nu$ is the bandwidth, and $t_\mathrm{samp}$ is the sampling time. In contrast to the approach used for estimating the RMS noise in our previous work \citep{abb+19b}, we do not directly compare the measured S/N for each pulsar with its catalogued flux density. This is to avoid underestimating the sensitivity in the presence of RFI excision algorithms which cause bright pulsars to be detected with reduced S/N or overestimating it for pulsars for which CHIME/FRB is sensitive only to the tail of the flux distribution and the catalogued flux density is not representative of that observed.

For each pulsar, we take the median of the measured S/N values on each day and account for the pulse width to estimate the ratio of the RMS noise at the pulsar sky location and its flux density. Assuming that the flux distribution for each pulsar is stable in time, the daily variation in this ratio for each pulsar is due to the varying RMS noise for its sky location. For each day, we normalize this ratio by dividing by its median over all days the pulsar was detected. We then perform a weighted average of the normalized ratios for all pulsars detected on each day to get the overall variation in RMS noise. Performing a weighted average ensures that pulsars with less intrinsic variation dominate the resulting estimate.

For (ii), we use a beam model to estimate sensitivity variation across a single transit of FRB 121102 (see Fig.~\ref{fig:localization}). The frequency-dependent model includes an analytic description of the FFT-formed beams \citep{nvp+17}, as well as an approximated forward-gain description of the primary beam, which is based on ray-tracing simulations done for the CHIME Pathfinder \citep{baa+15}. The composite model is used to compute the band-integrated relative sensitivity between different locations within the transit.

For (iii), we recognize the significant effect of our bandpass on detectability of the variety of spectra seen in FRB 121102 bursts.  To test our sensitivity to bursts with different spectral energy distributions (SED), we use the beam-former-to-Jansky conversion as a function of frequency that is generated from the calibration process. Multiplying this array with a simulated SED and summing across frequencies gives the expected recovered signal. We restrict this simulation to Gaussian spectral profiles, with which we convolve the conversion array to get relative S/N scale factors for different emission bandwidths and central frequencies.

We combine these three sources of variability in a Monte Carlo simulation, with each sample representing the fluence threshold for some possible burst within the defined exposure.  To generate a sample, we first draw a date within the covered interval, where the probability for choosing each day is proportional to its exposure (see upper panel Figure~\ref{fig:exposure}). We then draw from a Gaussian distribution parameterized according to the relative sensitivity and uncertainty for the chosen day, as given by the pulsar study (see lower panel Figure~\ref{fig:exposure}).  We then uniformly draw a position along the transit and within the FWHM at 600~MHz for the relevant beams (see Figure~\ref{fig:localization}). Since our decreased exposure is caused by a node failure, the active beams, and therefore possible positions, depends on the chosen day. Once a location is chosen, we compare sensitivity to the detection location to get our second relative sensitivity factor.

Next, we uniformly draw an emission bandwidth $\Delta\nu \in [\Delta\nu_{\mathrm{fit}}/2,\ 2\Delta\nu_{\mathrm{fit}}]$. The bandwidth determines the range of central frequencies we consider: $\nu \in [400\ \mathrm{MHz} + \Delta\nu/2,\ 800\ \mathrm{MHz} - \Delta\nu/2]$. The expected S/N for this SED is divided by the expected S/N for the detection-parameterized SED to get the final relative sensitivity factor ($\nu_\mathrm{fit} = 631$ MHz, $\Delta\nu_\mathrm{fit} = 100$ MHz; see Figure~\ref{fig:waterfall}).

The three factors are multiplied to get an overall relative sensitivity for the simulated sample. To translate this to a fluence threshold, we first draw an initial fluence threshold from a Gaussian distribution, which is parameterized according to the recovered S/N and measured fluence of our single detection.  Scaling the fluence to a detection threshold based on the S/N assumes a linear relationship between the two, which we caution is not always a valid assumption.  Finally, we divide the drawn initial fluence threshold by the overall relative sensitivity to get a fluence threshold for the simulated sample. We repeat this process a million times to build up a distribution of fluence thresholds. The median value with two-sided 90\% confidence interval is 7$^{+ 7}_{- 4}$ Jy~ms.  The medians and 90\% confidence intervals for the three intermediate distributions of relative sensitivity are (i) $1 \pm 0.2$, (ii) $0.7 \pm 0.3$, and (iii) $0.9 \pm 0.2$.

\subsection{400--800 MHz Rate}
\label{sec:rate}

Using the total estimated CHIME/FRB exposure time of 11.3 hrs (see \S\ref{sec:obs}) to FRB~121102 and the 90\% Poisson uncertainty on the single burst detected by CHIME/FRB of 0.05--4.7 bursts, we estimate an average rate in the CHIME band of 0.1--10 bursts per day for the interval 2018 July 25 through 2019 February 25.  We note that the repetition of FRB~121102 is known to be non-Poissonian \citep[e.g.][]{ssh+16b,oyp18} so this rate should be considered a rough approximation that is averaged over periods of likely variable activity.

\section{Discussion}
\label{sec:discussion}

We have presented the detection of a single burst from the first known repeater, FRB 121102, with CHIME/FRB and CHIME/Pulsar.  The burst has interesting morphology which we have characterized above, and which we now consider in light of proposed models for the progenitor and emission mechanisms.

Our detection represents the lowest radio frequency yet at which the source has been detected.  This demonstrates that any low-frequency cutoff frequency \citep[e.g.][]{rl19}, due, for example, to possible free-free absorption at the source, must be below $\sim$720~MHz, the lowest frequency at which we have observed signal, corrected for redshift. \citet{tbc+17} estimated the ionized gas properties in the host, predicting free-free absorption to be negligible even at 100 MHz, consistent with our observation.

Our analysis of the morphologically complex burst profile (see \S\ref{sec:morphology}), suggests a DM that is $\sim$1\% higher than that previously reported. Specifically, we find for this 2018 November burst, using a structure maximization method, DM = $563.6\pm0.5$~pc~cm$^{-3}$. We note that a contemporaneous measurement at 1.4 GHz finds a similar value (Seymour et al., in prep.).  Our numbers can be compared with the lower value of  560.5~pc~cm$^{-3}$ from \citet{hss+18} in 2016 September.  

So large an increase in DM is inconsistent with enhancements in the Milky Way electron column depth, as inferred from the much smaller DM changes seen in Galactic radio pulsars \citep[e.g.][]{pkj+13}.  Similarly, the IGM DM contribution seems unlikely to be so variable \citep{yz17}.  It is therefore most likely that the change is local to the source. For models that postulate FRB 121102 to be a young compact object inside a supernova remnant \citep{pg18}, the timescale for an increase in DM following the expected initial decrease depends on the nature of the progenitor star and the ambient density. We note that, in general, the expected rates of increase predicted by \citet{pg18} are significantly smaller than what we have observed.
\citet{mms19}, in their synchrotron maser model, predict stochastic DM variations from within
the remnant, a result of temporary increases from bursts that follow major flares from the central engine (in their case, a young magnetar). Due to Compton scattering however, they argue that these high DM events should be detectable preferentially at higher radio frequencies. This is at odds with
our lower frequency measurement, unless contemporaneous higher frequency measurements find an even higher DM.  On the other hand, an apparently increasing DM could be a result of intervening ejecta clumps passing through our line-of-sight; if so, a subsequent decrease of similar magnitude and time scale is expected.

Evidence for spectral flattening (not for individual bursts but for the overall burst rate --- a ``statistical'' spectral index) below 1~GHz has been reported for FRB 121102 \citep{hst+19,gms+19}. However, due to large uncertainties in the FRB 121102 burst luminosity function power-law index \citep{lab+17, gms+19} as well as the unclear and variable completeness of past searches for bursts from FRB~121102 \citep[see][]{gms+19}, we do not attempt to compare rates between samples with differing fluence thresholds. Such a comparison should be done when a large, uniform, complete sample of bursts from 1.4\,GHz is available. Furthermore, the comparison should be done on samples that are contemporaneous or using a framework that takes into account the variable burst rate \citep[e.g.][]{oyp18}, as the burst arrival times of FRB~121102 are known to be non-Poissonian \citep{ssh+16a,ssh+16b}.

Burst morphology is an important aspect to consider for both emission mechanism modelling and burst detection. Bursts with peaked spectral energy distributions are common, with FRB 121102 exhibiting emission bandwidths that appear to be proportional to frequency. The seven band-limited bursts at 3 GHz from \citet{lab+17} have average fractional bandwidth $\Delta\nu/\nu = 0.16 \pm 0.04$, the 18 band-limited bursts at 1.4 GHz from \citet{gms+19} have average fractional bandwidth $\Delta\nu/\nu = 0.12 \pm 0.04$, and the lone CHIME/FRB detection at 600 MHz has $\Delta\nu/\nu \simeq 0.15$. If similar fractional bandwidths also occur at lower frequencies, at 150 MHz we expect burst envelopes with FWHM of 20--30 MHz. Large bandwidth surveys operating at low frequencies may benefit from sub-band searches tuned to these expected burst bandwidths.

The sub-burst structure in time is also relevant for detection efforts.  Depending on the number of sub-bursts, the S/N-versus-DM curve can peak substantially beyond reference values obtained with structure-optimizing methods. Furthermore, the curve will flatten as the number of sub-bursts and envelope width increases. As flat curves are a characteristic feature of narrow-band RFI, this effect has direct implications for targeted searches that limit DM trials and classify candidates based on S/N behaviour in the DM-time plane.  Very aggressive RFI excision may be preferable if the search output is heavily contaminated.

The measured linear drift rate of $-3.9\pm0.2$ MHz/ms is consistent with the CHIME/FRB drift rates measured for the only other reported repeater, FRB 180814.J0422+73, of $\sim -1$ and $-6$ MHz/ms \citep{abb+19b}. Such a burst train has low-brightness-temperature analogs in the Sun, flare stars and planets \citep[see][and references therein]{hss+18} and might be explained as originating from an emitting region with a gradient in plasma conditions, e.g., in the deceleration of an FRB-producing blast wave \citep{mms19} or in the propagation from high- to low-curvature regions of bunches of charged particles \citep{wzc19}.  These models generally do not explicitly justify the seemingly disconnected sub-bursts in downward drifting bursts yet, but derive a relation $\dot{\nu}_\mathrm{obs} \propto \nu_\mathrm{obs}^\alpha$, with $\alpha$ depending on the specifics of the progenitor model.

\begin{figure}[h]
	\begin{center}
		\includegraphics[scale=0.8]{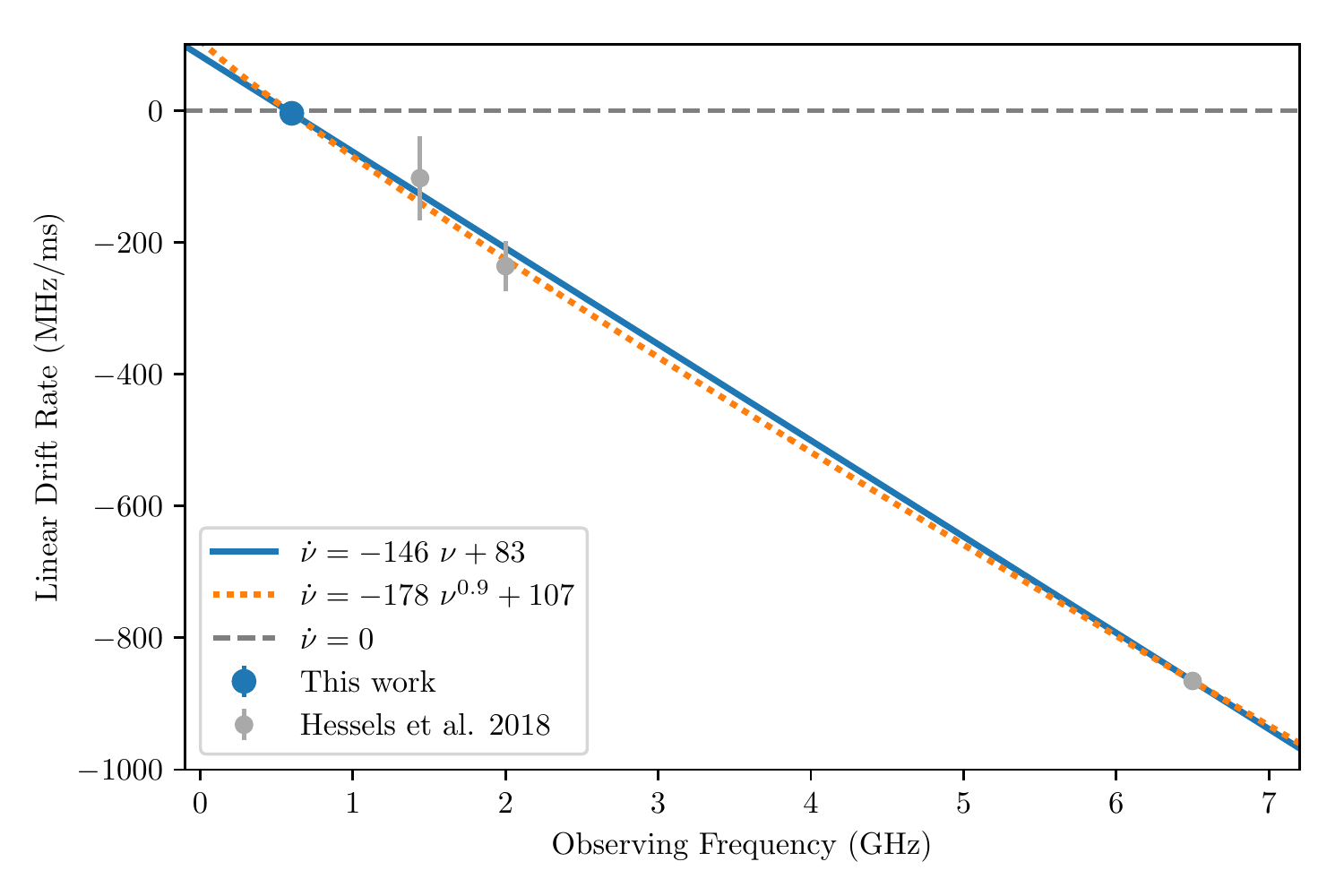} %{R1_drift.pdf}
	\end{center}
\figcaption{Comparison of FRB 121102's CHIME/FRB burst drift rate with rates measured by \citet{hss+18} at higher observing frequencies. Where multiple bursts are detected with the same instrument (at 1.4 and 2.0 GHz), the weighted average and standard deviation are taken at the central frequencies of the receiver bandwidth and error bars reflect the scatter in the measurements, while the error bars at 600 MHz and 6.5 GHz reflect the measurement errors. Over this frequency range, the drift rate evolves roughly linearly with $\sim -$150 MHz/ms/GHz.}
\label{fig:driftrates}
\end{figure}

In Figure~\ref{fig:driftrates}, we compare our measured FRB 121102 burst drift rate and rates for events from this source measured at 1.4--6.5 GHz. A linear drift rate evolution ($\alpha=1$) of $\sim -$150 MHz/ms/GHz with an offset of $\sim$80 MHz/ms fits the observed drift rates well, but no direct relation between $\dot{\nu}$ and $\nu$ can be drawn. Leaving the power-law index free, the data favor $\alpha \sim 1$ with similar slope and intersection as the fit with fixed index. Assuming burst components are extended as has so far been observed (order 100 MHz and 1 ms) the maximum observable linear drift rate for a given data set is approximately the frequency bandwidth BW divided by the time resolution $t_\mathrm{s}$, while the minimum observable drift rate is dominated by the frequency channel bandwidth $\Delta f_\mathrm{c}$ (although besides component widths, the sensitivity to drift rates also largely depends on the number of components and their S/N). CHIME/FRB (BW=400 MHz; $t_\mathrm{s}\sim1$ ms) is thus sensitive to linear drift rates up to $\sim-400$ MHz/ms and from the Arecibo and GBT data sets ($\Delta f_\mathrm{c}\sim 1$ MHz) drift rates as low as $\sim-1$ MHz/ms would likely be measurable --- making all observations sensitive to a wide range of linear drift rates. If the apparent linear trend in Figure~\ref{fig:driftrates} were to hold down to the LOFAR band, the drifting rate would naively be positive, +61 MHz/ms at 150 MHz.  This would be difficult to observe if bursts have small fractional bandwidths as discussed above, and certainly if bursts are heavily scattered as expected in the LOFAR band.

The upper limit on the scattering time at 500 MHz (\S\ref{sec:morphology}) in our burst detection implies a $3\sigma$ upper limit of 0.6 ms at 1 GHz, which is unremarkable in the FRB population \citep[see e.g.][]{abb+19a}.  Indeed this upper limit is significantly higher than the scattering time measurements we have made for other CHIME/FRB events. Relevant here is sub-burst structure and the overall width of the FRB 121102 burst we have detected, 34 ms, much greater than those of our other reported FRBs, for which much shorter scattering times were measurable.  In any case, our upper limit at 1 GHz is two orders of magnitude higher than the scattering time at 1 GHz (24~$\mu$s) inferred from the scintillation bandwidth at 5 GHz \citep{msh+18}, and provides no evidence for a second screen as observed, e.g., by \citet{mls+15a}.

\section{Conclusions}

We have detected a single burst from the original repeating FRB 121102 using the commissioning CHIME/FRB system. Our detection represents the lowest radio frequency at which the source has yet been detected, $\sim$600 MHz.  The burst has complex morphology and shows no evidence for scattering, but tentatively appears to be at a DM that is $\sim$1\% higher than has previously been measured. The sub-bursts show downward drifting in radio frequency that has now been observed in many FRB 121102 bursts as well as in a second repeating source \citep{abb+19b}.  Considering previously reported drift rates in FRB 121102 bursts at different frequencies, we note a possible linear trend in drift rate as a function of observing frequency, similar to that expected in multiple emission models. Our estimated average burst rate for FRB 121102 during our pre-commissioning and commissioning phases based on this single burst is 0.1--10 per day, which is consistent with rates reported at higher radio frequencies, although we recognize the importance of comparing contemporaneous rates since the source rate varies.  Fortunately, CHIME/FRB observes the source position daily so ultimately should be useful for comparing CHIME-band rates with those measured simultaneously at higher frequencies.

\acknowledgements

We are grateful to the staff of the Dominion Radio Astrophysical Observatory, which is operated by the National Research Council Canada.
The CHIME/FRB Project is funded by a grant from the Canada Foundation for Innovation 2015 Innovation Fund (Project 33213), as well as by the Provinces of British Columbia and Quebec.
Additional support was provided by the Canadian Institute for Advanced Research (CIFAR) Gravity \& Extreme Universe Program, McGill University and the McGill Space Institute, University of British Columbia, and University of Toronto Dunlap Institute.
The Dunlap Institute is funded by an endowment established by the David Dunlap family and the University of Toronto.
Research at Perimeter Institute is supported by the Government of Canada through Industry Canada and by the Province of Ontario through the Ministry of Research \& Innovation.
Research at the McGill Space Institute is supported in part by a gift from the Trottier Family Foundation.
The National Radio Astronomy Observatory is a facility of the National Science Foundation operated under cooperative agreement by Associated Universities, Inc.
P.C. is supported by an FRQNT Doctoral Research Award and a Mitacs Globalink Graduate Fellowship.
B.M.G. acknowledges the support of the Natural Sciences and Engineering Research Council of Canada (NSERC) through grant RGPIN-2015-05948, and of the Canada Research Chairs program.
V.M.K. holds the Lorne Trottier Chair in Astrophysics \& Cosmology, a Canada Research Chair, and the R. Howard Webster Foundation Fellowship of CIFAR, and receives support from an NSERC Discovery Grant and Herzberg Award, and from the FRQNT Centre de Recherche en Astrophysique du Qu\'ebec.
M.M. is supported by a NSERC Canada Graduate Scholarship.
Z.P. is supported by a Schulich Graduate Fellowship.
S.M.R. is a CIFAR Senior Fellow and is supported by the NSF Physics Frontiers Center award 1430284.
P.S. is supported by a DRAO Covington Fellowship from the National Research Council Canada.
The baseband system is funded in part by a CFI JELF award to I.H.S.

%\software{alpenhorn, presto}

\bibliographystyle{aasjournal}
\bibliography{frbrefs}

\end{document}